\documentclass[journal]{IEEEtran}
\usepackage[noadjust]{cite}

\ifCLASSINFOpdf
\else
\fi
\usepackage{amsmath}
\usepackage{mathrsfs}
\usepackage{bm}
\usepackage{amssymb}

\usepackage{graphics}
\usepackage{makecell}
\usepackage{graphicx}
\usepackage{enumerate}

\begin{document}


%
%
%
\title{Controlling Power and Virtual Inertia from Storage for Frequency Response}
%
%
%


\author{Shuchang~Yan,~\IEEEmembership{Student Member,~IEEE}
}

\maketitle

\begin{abstract}
Nowadays, power imbalance happens more frequently due to the more integration of renewable energy sources. Energy storage is a kind of devices that can charge energy at one time and discharge energy at another time. This function makes that storage is widely envolved into promoting power balance of power system. Besides this function, storage can also emulate virtual inertia to respond to frequency deviations in the system. This work provides a generalized optimization framework
to analyze how to control power and virtual inertia from storage to participate in frequency response when a large disturbance happens. Centralized and distributed model predictive control is employed here, and case study verifies the effectiveness of our optimization framework.

\end{abstract}

\begin{IEEEkeywords}
energy storage, power, virtual inertia, model preidctive control
\end{IEEEkeywords}

\IEEEpeerreviewmaketitle

\section{Introduction}


Wind power and solar power have been brought into power system considering that we want to build a more enviromentally and economically friendly society, and this makes that there appears high ramping for the netload in power system~\cite{ayodele2015mitigation}. One of the functions for storage is to smooth the load and further reduce power mismatch, which makes it easy for conventional generators to catch up. Power imbalance will cause frequency deviations, and it is not desired for the equipments in power system. Another fact is that more and more conventional generators have been replaced by renewables~\cite{ulbig2014impact}, frequency problems will be more severe. To face these challenges, researchers propose the concept and develop the technique ``virtual inertia" to secure the frequency of power system for power electronic deivices including energy storage~\cite{fang2018inertia}. Until now, there are two functions for storage to deal with power imbalance in system, virtual inertia and providing charged/discharged power. And the research question is 
{\itshape{how we control power or virtual inertia from storage  for frequency response?}}

To answer this question, the characteristics of storage are reviewed first. Storage have different categories including chemical, thermal, electrical, electrochemical and mechanical storage~\cite{zzz3}. Because we focus on applying energy storage into frequency control of power system, electrochemical storage (or battery storage) will be the primary task takers for its fast acting speed. Normally, there will be only tens to hundreds of milliseconds for battery storage to reach its power capacity~\cite{hsieh2012frequency,delille2012dynamic}. This characteristic makes that storage can provide charged/discharged power to the system when power imbalance happens, which can reduce the frequency deviations. And this method can be applied in primary frequency control~\cite{wen2016frequency} and security constrained OPF~\cite{wen2015enhanced}.

 On the other hand, storage, as a kind of power-electronic devices, can implement virtual inertia to respond to frequency deviations~\cite{kerdphol2018demonstration}. And virtual inertia can be fixed with time (time-constant) or changes with time (time-variant)~\cite{tielens2016relevance}. Traditionally, virtual inertia, as parameters in control process (mainly represented by transfer function), is fixed~\cite{kiaei2018tube, rostami2018scalable}. To analyze the effects of virtual inertia, the optimization objective functions include minimizing the angle differences ~\cite{kiaei2018tube, rostami2018scalable,adrees2016study} or minimizing the costs from power electronic devices~\cite{guggilam2018optimizing}. For the time-variant virtual inertia, there is only one work addressing that how to control time-variant virtual inertia from storage can have desired frequency trajectory~\cite{
yan2019controlling}, but there still remains a open question that in frequency regulation process, which is better between controlling power or virtual inertia from storage to achieve a certain control objective. 

To optimize the desired index (stability or economic) for power system, myopic optmization and looking ahead optimization are two major optimization techniques~\cite{messac2015optimization}. Myopic optimization means optimization is conducted when all the information is known. However, in power engineering practice, there are many uncertain resources such as loads and wind power. Many forecast methods have been developed to help us to get better command of these uncertain parameters~\cite{lei2009review, senjyu2002one}, but absolute accuracy can not be reached. To better utilize these forecast information, looking ahead optimzation technique (mainly represented by model predictive control (MPC)~\cite{rawlings2009model}) is widely applied. And also, distributed MPC is also developed and applied to meet the needs that decisions should be made by different control centers at different places of power system~\cite{camponogara2002distributed, venkat2008distributed}. 

Charged/discharged power and virtual inertia are two kinds of control resources (or decision variables) for storage, and they have different units.  In this work, we aim to give a optimization framework to analyze how to optimally dispatch these two control resources. 
The objective is to minimize frequency deviations of all buses in the system. The constraints include the frequency contraints at certain buses and the power/energy constraints of storage. This optimization framework is utilized to do the following analysis:

 \begin{enumerate}[i.]
	\item This optimization framework can give a decision that how to allocate virtual inertia and dispatch power to meet a certain control objective. 
	\item This optimization framework can make the resources allocation decisions to be determined in a distributed way, which can meet the needs of current power system with multiple control centers.
	\item Model predictive control is employed to incorporate the potential forecast errors from uncertain resources such as wind power and load power.
\end{enumerate}

The remainder of this work is structured as follows. In Section II, the structure preserving model and energy storage model are presented. Section III presents the mathematical model of this optimization framework and illustrates how the distributed model predictive control works. Centralized MPC is conducted on a two-bus system and distributed MPC is conducted on a 12-bus system in Section IV. The last section, Section V, gives the final conclusion.

\section{Modeling}

\subsection{Power System Dynamics}

The structure preserving model~\cite{bergen1981structure} is utilized here to model the dynamics of power system. In this model, reactive power is ignored and voltage magnititudes at all buses are assumed to be constant at 1 per unit value (p.u.). And all the transmission lines in power system are assumed to be resistanceless. There are $N$ nodes or buses in power system. Nodes or buses with inertia and without inertia are denoted as the set $G$ and set $L$ respectively, and superscripts $g$ and $l$ are the elements from $G$ and $L$ respectively. In power system, nodes with inertia are generally buses with generators or motors, and the corresponding dynamics are described as follows, 
\begin{align}
\dot{\delta_{i}^{g}}=\omega_{i}^{g} \label{eq1}\\
\dot{\omega}_{i}^{g}=-\frac{D_{i}}{M_{i}}\omega_{i}^{g}-\frac{1}{M_{i}}&\sum^{N}_{\substack{j=1 \\ j  \neq i}}b_{ij}\sin(\delta_{i}-\delta_{j})+\frac{1}{M_{i}}P_{i}^{0} \label{eq2}
\end{align}
where $\delta_{i}$ and $\delta_{j}$ are the angle of bus $i$ and bus $j$ respectively, and $M_{i}$ and $D_{i}$ are the inertia and damping of bus $i$ respectively, $\omega_{i}$ is the frequency  at bus $i$, and it takes the nominal frequency as a reference. $b_{ij}$ is the susceptance between bus $i$ and bus $j$, and $b_{ij}\sin(\delta_{i}-\delta_{j})$ is the active power flow from bus $i$ to bus $j$, $P_{i}^{0}$ is a shorthand for $P_{M,i}^{0}-P_{D,i}^{0}$, which is the difference between mechanical power input $P_{M,i}^{0}$ of generator at bus $i$ and load demand $P_{D,i}^{0}$ at bus $i$. For the nodes without inertia, ususally load buses without motor loads, the dynamics can be descibed as follows,
\begin{align}
\dot{\delta}_{i}^{l}=-\frac{1}{D_{i}}&\sum^{N}_{\substack{j=1 \\ j  \neq i}}b_{ij}\sin(\delta_{i}-\delta_{j})+\frac{1}{D_{i}}P_{i}^{0} \label{eq3}
\end{align}
where the mechanical power input $P_{M,i}^{0}$ at buses without inertia is equal to 0 and $P_{i}^{0}$ is equal to $-P_{D,i}^{0}$, representing the active power drawn from the node $i$.

\subsection{Energy Storage Dynamics}

Energy storage can mimic the behaviour of synchronous generators to provide virtual inertia and damping, the dynamics of energy storage are given below~\cite{guggilam2018optimizing},
\begin{align}
\dot{\delta_{i}^{e}}=\omega_{i}^{e} \label{eq4}\\
\dot{\omega}_{i}^{e}=-\frac{D_{e, i}}{M_{e,i}}\omega_{i}^{e}-\frac{1}{M_{e, i}}&\sum^{N}_{\substack{j=1 \\ j  \neq i}}b_{ij}\sin(\delta_{i}-\delta_{j})+\frac{1}{M_{e, i}}P_{i}^{e} \label{eq5}
\end{align}
where $M_{e, i}$ and $D_{e, i}$ are virtual inertia and damping for energy storage at bus $i$ respectively, $P_{i}^{e}$ is the constant or reference power input or power output for energy storage at bus $i$. The nodes or buses with energy storage are denoted with the set $S$. For simplicity's consideration, there exists $N=G\cup L \cup S$, $ G \cap L=\varnothing$, $ G \cap S=\varnothing$ and $ L \cap S=\varnothing$.

\subsection{Constraints for the System Dynamics}

The dynamics of power system with energy storage can be expressed by~(\ref{eq1})-(\ref{eq5}). However, there are some practical limits such as the frequency limits for several certain buses. The system constraints are listed as follows,
\begin{gather}
|\omega_{i}| \leq \omega_{i}^{max}\\
P_{i}^{e,min}\leq P_{i}^{e} \leq P_{i}^{e,max}\\
\label{eq8}
E^{al,l}_{i}\leq \int_{t_{0}}^{t_{1}}P_{i}^{e} dt\leq E^{al,u}_{i}
\end{gather}
where $\omega_{i}^{max}$ is the allowable frequency change of bus $i$, $P_{i}^{e,min}$ and $P_{i}^{e,max}$ are the lower and upper limits of power for energy storage at bus $i$, and $E^{al,l}_{i}$ and $E^{al, u}_{i}$ are the lower and upper limits of energy change for storage at bus $i$, and $t_{0}$ and $t_{1}$ are the initial and final time instance of concerned time interval.

\section{Problem Formulation}
In this section, the formulation for optimization problem will be given. Centralized MPC will be reviewed first and the procedure to implement distributed MPC will be introduced. 
\subsection{Centralized MPC}

MPC calculates the optimal control inputs for a certain objective at a selected future time horizon, and then only the control input at the first time step is implemented. And this process is repeated until the concerned full time period is reached. This control method can better utilize future/predicted information and feedback advantage is included.

\vskip 0.3cm

\begin{gather}
\label{eq24}
\underbrace{\sum_{i=1}^{K}(\sum_{s=1}^{S}c_{p,s}P_{s}^{e}(k)/P^{b}+\sum_{s=1}^{S}c_{m,s}M_{e,s}(k)/M^{b})\times{T_{s}}}_{control  ~~effort}\nonumber \\
\underbrace{+\sum_{i=1}^{K}(\sum_{i=1}^{N}|{\omega_{i}|})\times{T_{s}}}_{control~~performance}
\end{gather}

subject to
\begin{gather}
System~dynamic~constraints~~~(1)-(5)\nonumber\\
~~Frequency~contraints~~~(6)\nonumber\\
~~Power/Energy~contraints~~~(7),~(8)\nonumber
\end{gather}
where $P_{s}^{e}(k)$ is the reference charged/discharged power of storage and $M_{e,s}(k)$ is virtual inertia provided by storage at bus $s$, $c_{p,s}$ and $c_{m,s}$ are the corresponding cost coefficients for power and virtual inertia respectively, $K$ is the total time steps in the control horizon.

It is noted that there are two objectives in the cost function, and they have different units. To compare the perfermances with different weights, we eliminate the units of $P_{s}^{r}(k)$ and $M_{e,s}(k)$ by dividing the base value $P^{b}$ and $M_{}^{b}$ respectively.  In the meanwhile, we need to discretize the system dynamics described by $(\ref{eq1})$-$(\ref{eq8})$, Euler discretization method is utlized~\cite{bayer2006discretization}, and $T_{s}$ is the discretization time step.
\subsection{Distributed MPC}

Distributed MPC calculates the objective function in centralized MPC with a distributed way~\cite{christofides2013distributed}. Distributed calculation allows different control centers in power system to make the decision independently with limited information exchange, and distributed MPC has been applied in many aspects of power system~\cite{morstyn2018model,hans2019hierarchical}. 

To compute the objective function in a distributed way, decompostion method is needed. In this work, we adopt the a proximal DC-ADMM (PDC-ADMM) method~\cite{chang2014multi,chang2016proximal}. We list the main process to implement the distributed MPC,

\vskip 0.15cm
Step 1)\quad{\itshape{Initial Scenario/Parameter Setting}}
\begin{itemize}
	\item[]
	\setlength\leftskip{0em} 
	
\quad Partition power system into different areas (1,2,...,$A$) without overlapping each other, and one control center is assumed to exist in one control center.

	\quad Choose the whole time interval for the simulation $T_{total}$, the time horzion for MPC $T_{h}$, and the discretization step $T_{s}$.
	
		\quad Determine the disturbance for the power system.

\end{itemize}

\vskip 0.15cm

Step 2)\quad{\itshape{Reformulate the centralized MPC}}
\begin{itemize}
	\item[]
	\quad We first reformulate the problem into the following forms,
	\begin{gather}
\label{eq25}
\sum_{a=1}^{A}(\sum_{p=1}^{Ap}c_{p,a}P_{p,a}^{e}(k)/P^{b}+\sum_{m=1}^{Am}c_{m,a}M_{m,a}(k)/M^{b})\times{T_{s}}
~\nonumber\\
+\sum_{a=1}^{A}\sum_{g=1}^{Ag}c_{g,a}{|\omega_{g,a}|}\times{T_{s}}\\
subject~to \nonumber~~~~~~~~~~~~~~~~~~~~~~~~~~~~~~~~~~~~~~~~~~~\\
G_{a}(X_{a})\leq 0,~~a=1,...,A\\
H_{a}(X1,...,X_{A})\leq 0,~~a=1,...,A
	\end{gather}
	Where $a$ is the index for the area set $A$, $A_{p}$ is the total number of energy storage whose reference power can be changed in area $a$, $A_{m}$ is the total number of energy storage whose virtual inertia can be changed in area $a$, $A_{g}$ is the total number of buses whose frequencies are concerned in area $a$, $G_{a}(X_{a})$ are the seperated constraints and $H_{a}(X_{1},...,X_{A})$ are the coupling constraints. 
	
	\quad And we denote the objective function in each area as $F_{a}(X_{a})$, and it can expressed as follows,
		\begin{gather}
	\label{eq26}
	F_{a}(X_{a})=(\sum_{p=1}^{Ap}c_{p,a}P_{p,a}^{e}(k)/P^{b}\nonumber \\+\sum_{m=1}^{Am}c_{m,a}M_{m,a}(k)/M^{b})\times{T_{s}}\nonumber \\
	+\sum_{g=1}^{Ag}c_{g,a}{|\omega_{g,a}|}\times{T_{s}}
	\end{gather}

\end{itemize}
\vskip 0.15cm
Step 3)\quad{\itshape{Conduct distributed MPC}}
\begin{itemize}
	\item[] 
	\setlength\leftskip{1em}
	\quad At the initial time horizon, we initialize the primal variables in each area and Lagragian multipliers for the coupling constraints, do one step simulation at each subproblem, update primal variables and Lagragian parameters as in {\bf Algorithm 2} in~\cite{chang2016proximal} .

	\quad We implement the control input at the first time step of the initial time horizon. When the system moves to next time step, we do the distributed optimization at the next time horizon.
	
	\quad The precedure in Step 3) continues until the time moves to the end of the whole time interval.
\end{itemize}

\section{Case Study}
In this section, we will implement centralized MPC and distributed MPC on 2-bus system and 12-bus system respectively. We will compare the performances in the cases where the frequency regulation is purely by virtual inertia or charged/discharged power of storage, and by combining  virtual inertia and charged/discharged power of storage. And we will consider energy constraints of energy storage, and the lower and upper constraints as -45 (p.u.$\cdot$s) and 10 (p.u.$\cdot$s) respectively.

\subsection{2-bus system}

Typical parameters will be utilized for this two bus system for simplicity's consideration, and centralized model predictive control is implemented. The inertia and damping of generator are fixed at 3s and -1 p.u. respectively during the transient process. The range of virtual inertia for storage at bus 2 is [1s, 15s], and the damping is -1 p.u. The  power output of generator at bus 1 is 3 p.u., power charge of storage at bus 2 is -3 p.u. at the initial timestep, and the $b_{i,j}=50$. There is 0.2 p.u. power increase at bus 1. The initial state of charge of storage is 0 p.u.$\cdot$s.

\begin{figure}[!tbh]
	\centering
	\includegraphics[width=2.5in]{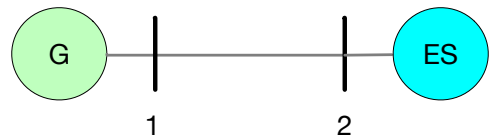}
	\caption{2-bus test system.}
	\label{n1}
\end{figure}

For MPC on this two-bus system, the discretization time step $T_{s}$=0.01s, and the looking-ahead time interval is $T_{h}$=0.1s.

\subsubsection{Constant Virtual inertia $\&$ Constant Power} 
\begin{figure}[!tbh]
	\centering
	\includegraphics[width=3.3in]{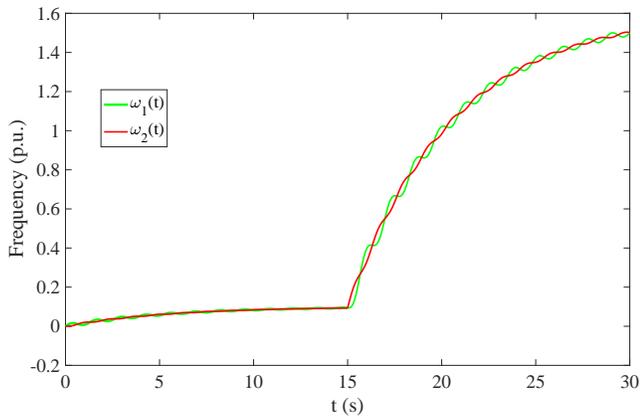}
	\caption{Frequencies for the two buses.}
	\label{n2}
\end{figure}

In this case, we fixed the power charge of storage at -3 p.u. and virtual inertia of storage at 8 s. And the frequencies of these two buses are given in~Fig.~\ref{n2}. For this case, the energy increase of storage is 90 (p.u.$\cdot$s) from $t$=0s to $t$=30s. 

It can be seen that energy stored at the storage has reached the lower limit at $t$=15s, and then the system frequency experiecne a large increase.

\subsubsection{Constant Virtual Inertia $\&$ Variant Power} 
\begin{figure}[!tbh]
	\centering
	\includegraphics[width=3.2in]{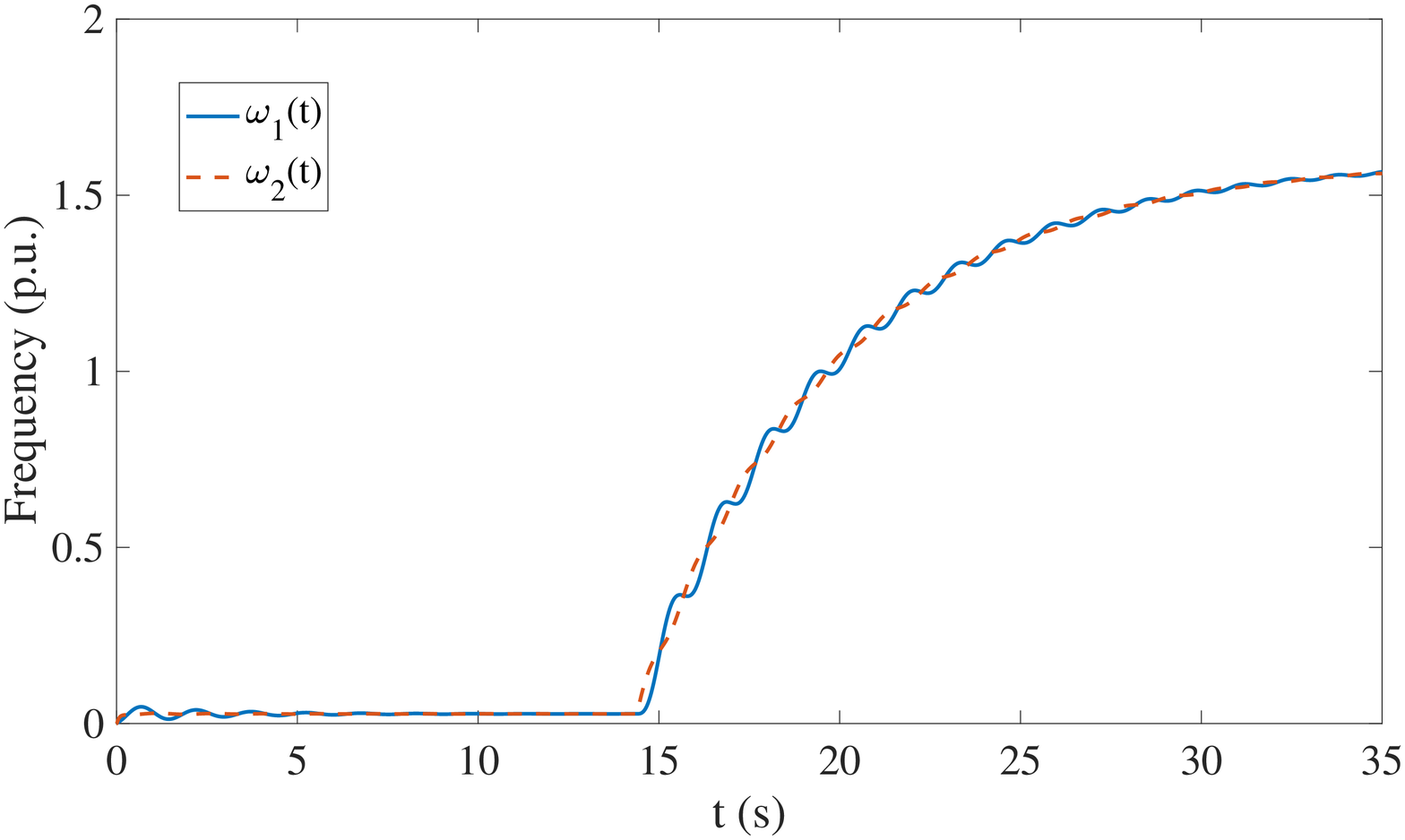}
	\caption{Frequencies of two buses.}
	\label{ne6}
\end{figure}

\begin{figure}[!tbh]
	\centering
	\includegraphics[width=3.2in]{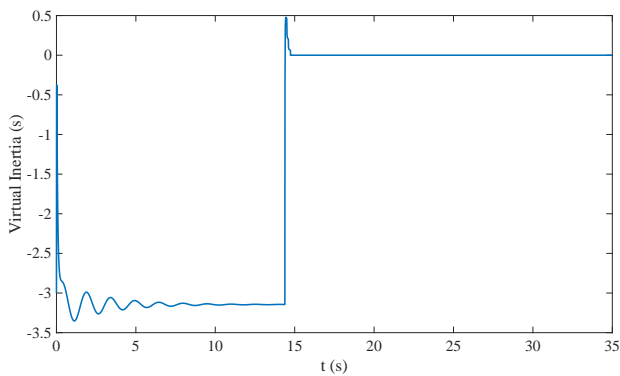}
	\caption{Power change  of storage.}
	\label{ne7}
\end{figure}

The frequency of these two buses, and power change of storage are shown in the Fig.~\ref{ne6} and Fig.~\ref{ne7} respectively.

\subsubsection{Variant Virtual Inertia $\&$ Constant Power} 
\begin{figure}[!tbh]
	\centering
	\includegraphics[width=3.2in]{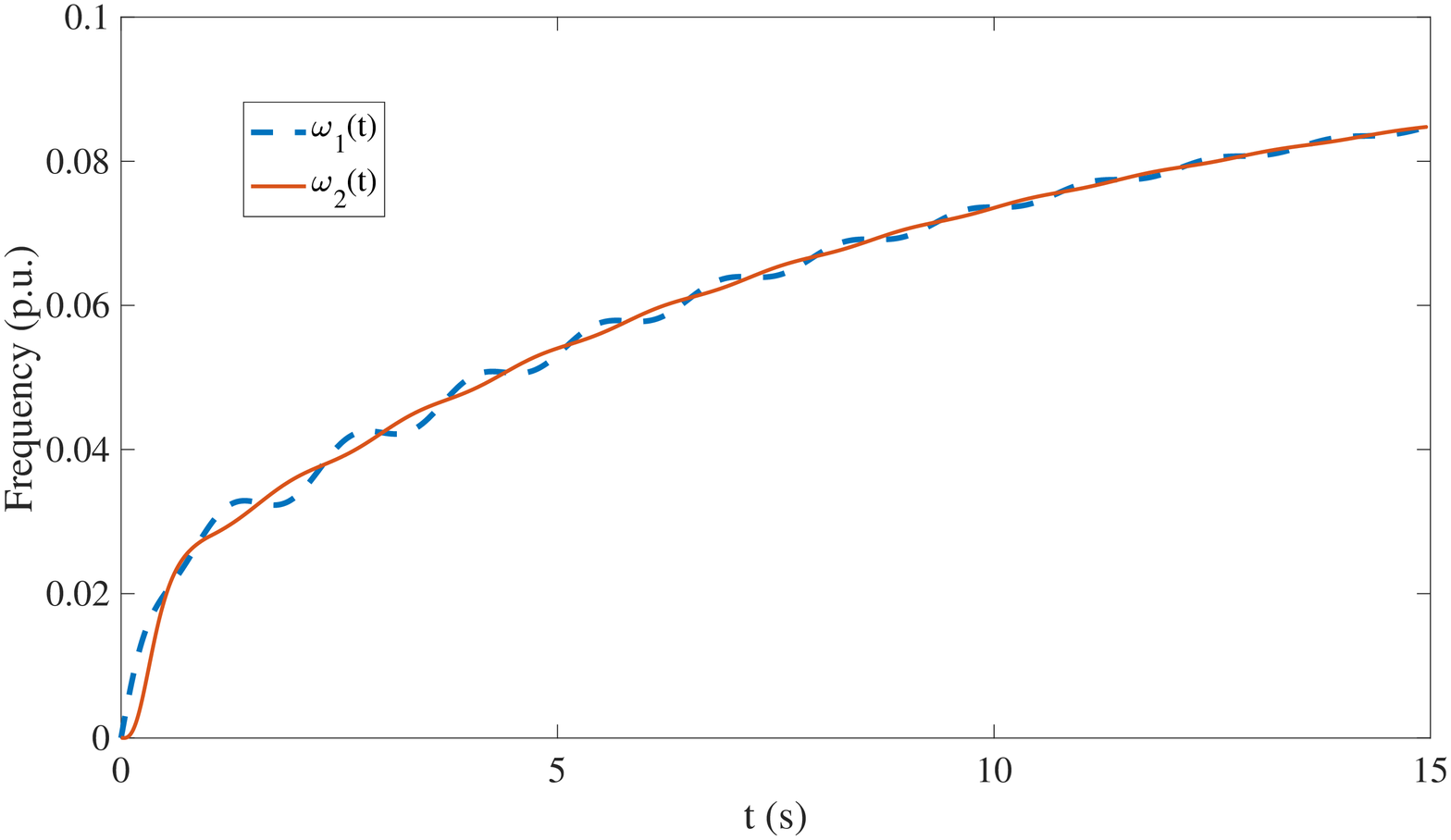}
	\caption{Frequencies of two buses.}
	\label{ne5}
\end{figure}

In this case, only the virtual inertia from storage can be changed. The frequnecy of these two buses are shown in Fig.5. And it can be seen that  there is no too much difference for the magnititudes of frequency deviations between Fig.~\ref{n2} and Fig.~\ref{ne5}.

\begin{figure}[!tbh]
	\centering
	\includegraphics[width=3.15in]{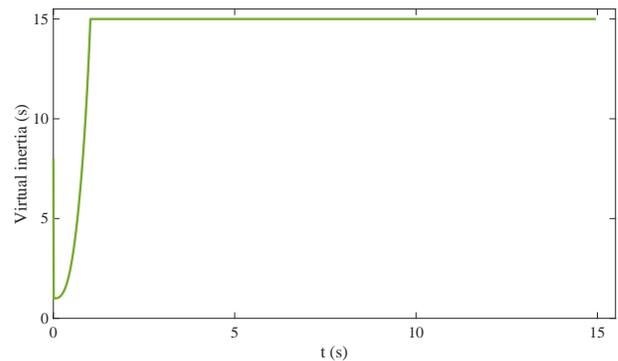}
	\caption{Virtual inertia change of storage.}
	\label{ne4}
\end{figure}

It also can be seen that the virtual inertia has a decrease at the initial stage and climbs to the maximum value (15s) for the following time to hinder the frequency increase. 

\subsubsection{Variant Virtual Inertia $\&$ Variant Power}
In this case, virtual inertia and power from storage can be changed, and the frequencies of this two-bus system are shown as follows, 

\begin{figure}[!tbh]
	\centering
	\includegraphics[width=3.2in]{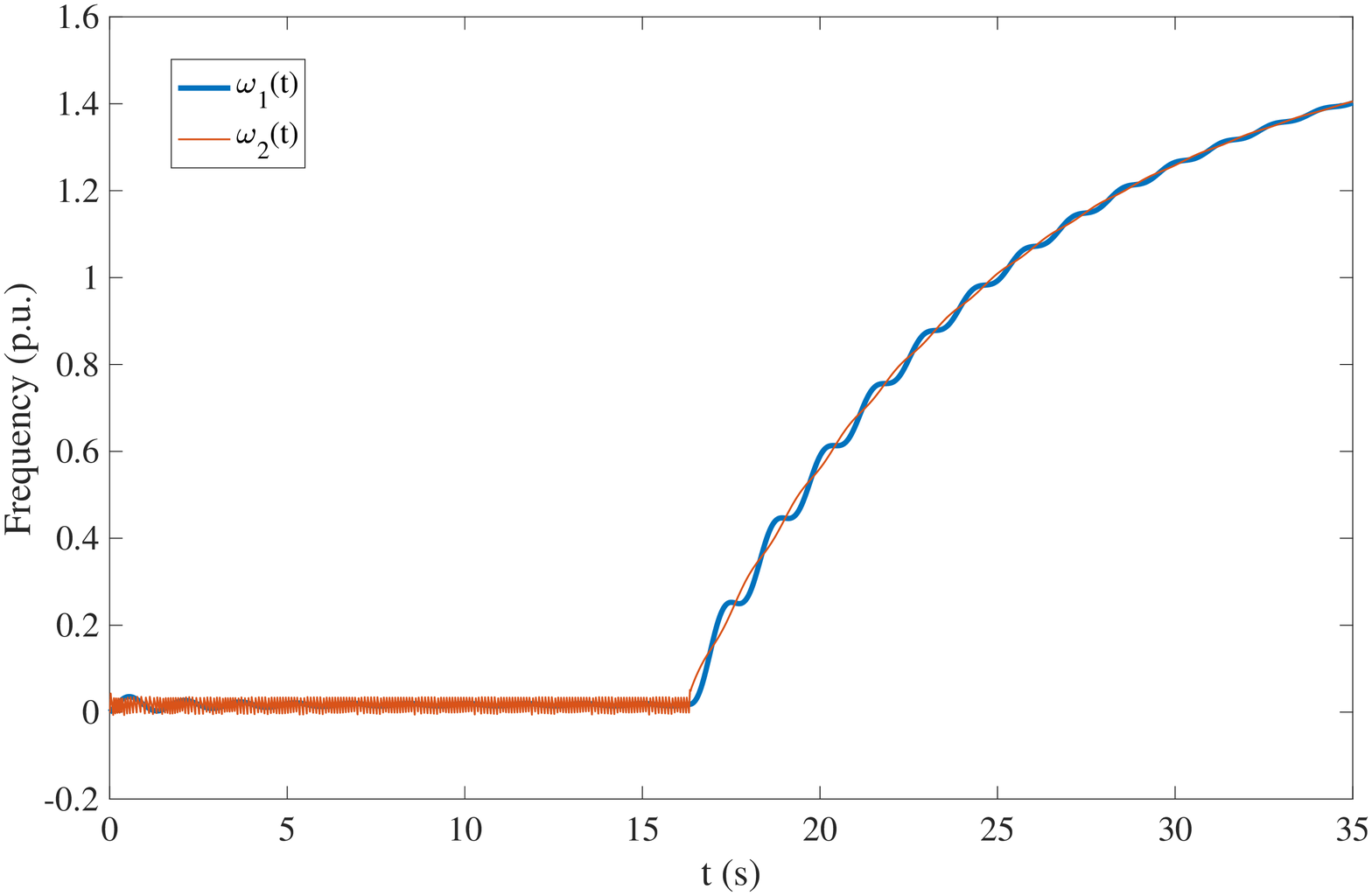}
	\caption{Frequencies of two buses.}
	\label{ne1}
\end{figure}

\begin{figure}[!tbh]
	\centering
	\includegraphics[width=3.3in]{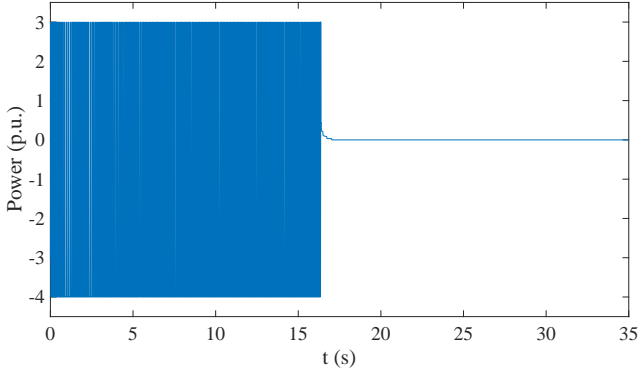}
	\caption{Power change of storage.}
	\label{ne2}
\end{figure}

\begin{figure}[!tbh]
	\centering
	\includegraphics[width=3.3in]{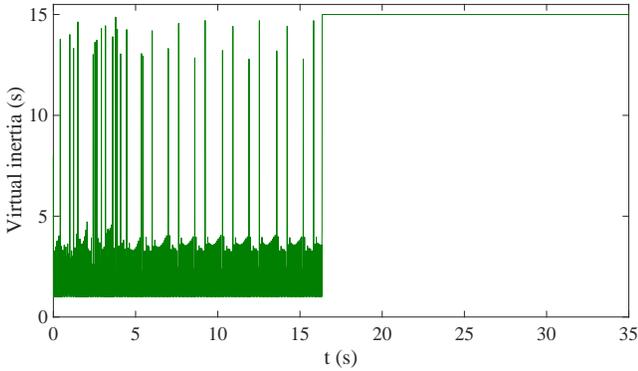}
	\caption{Virtual inertia change of storage.}
	\label{ne3}
\end{figure}

Comparing Fig.\ref{ne1} with Fig.\ref{n2}, it can be seen that when power and virtual inertia from storage can be controlled, the frequency change of these two buses has less deviations than the one where power and virtual inertia is fixed. However, when the energy from storage reaches the lower limit, controlling power is useless, and thus, virtual inertia is controlled at the maximim value (15s) to deter the increase of the frequency, as shown in Fig.\ref{ne2} and Fig.\ref{ne3}. 

In this case, we can see that only controlling virtual inertia of storage will significantly influence the frequecies of the two buses by comparing Fig.~\ref{ne5} and Fig.~\ref{n2}. Controlling power will have more effects on stabilizing the two-bus system frequency by comparing Fig.~\ref{ne6} and Fig.~\ref{n2}. This is reasonable because controlling power from storage can balance the power increase at bus 1, however, only controlling virtual inertia can not undertake this task.

\subsection{12-bus system}
\begin{figure}[!tbh]
	\centering
	\includegraphics[width=3.5in]{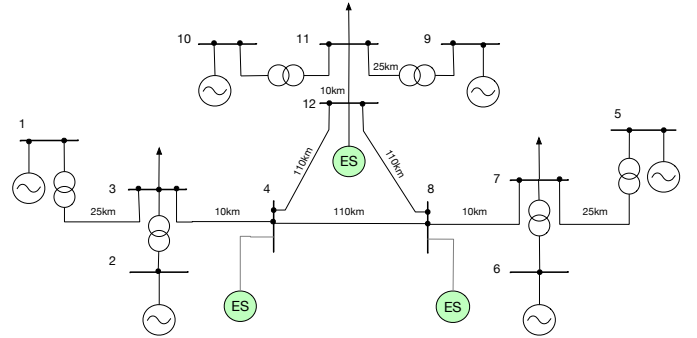}
	\caption{12-bus test system.}
	\label{f2}
\end{figure}

The 12-bus test system in Fig.~\ref{f2} is modified from the well-known two-area system in reference~\cite{kundur1994power} and an additional area is added as reference~\cite{borsche2015effects}. It contains 6 generators and 6 loads. The transformer reactance is 0.15 p.u. and the line impedance is (0.0001+0.001i) p.u./km. We still utilize structure preserving model to describe the dynamics of power system. The base capacity of this system for power flow calculation is set as 100MVA. The inertia and damping of original power system is given in Table~\ref{tab1} and the steady power flow condition is given in Table~\ref{tab22}.  It is assumed that there are motor loads (including little inertia and damping) at the load buses and bus 9 is a set as a reference bus in the system.

The time step is $T_{s}$=0.02s, the prediction horizon is $T_{h}$=0.12s the time interval for running the simulation is $T_{total}$=40s. 
The contingency setting is power increase of 20MW (0.2 p.u.) at bus 1. In this case study, we will both control power and virtual inertia from storage to regulate the system frequency. Area 1 includes buses 1, 2, 3 and 4, area 2 includes buses 5, 6, 7 and 8, and area 3 includes buses 9, 10, 11, and 12.

\begin{figure}[!tbh]
	\centering
	\includegraphics[width=3.5in]{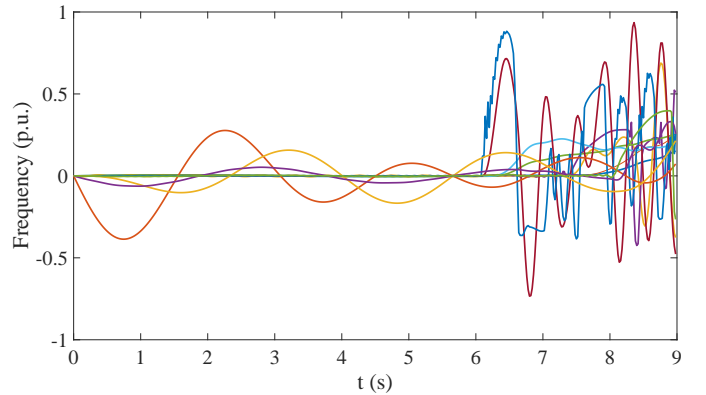}
	\caption{Frequencies of 12-bus system.}
	\label{nt1}
\end{figure}

The frequencies of all the buses in this system is shown in Fig.~\ref{nt1}, it can be seen that before $t$=5.5s, the frequencies tend to be stable, however, after $t$=5.5s the system frequency becomes unstable and large oscillations happens. 

\begin{figure}[!tbh]
	\centering
	\includegraphics[width=3.5in]{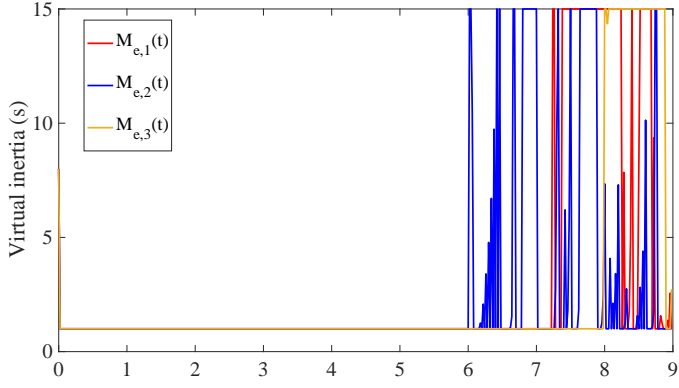}
	\caption{Virtual inertia change of storage.}
	\label{nt2}
\end{figure}

Figure~\ref{nt2} shows that virtual inertia all choose to be the minimum value 1s before $t$=6s, and after this, virtual inertia change significantly to regulate the fluctuating frequency. 

\begin{figure}[!tbh]
	\centering
	\includegraphics[width=3.5in]{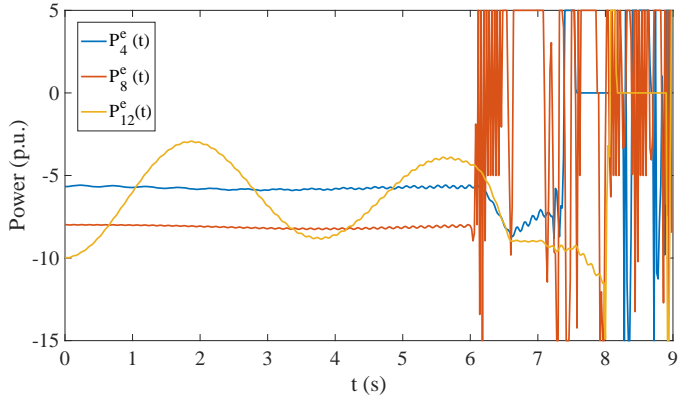}
	\caption{Power change of storage.}
	\label{nt3}
\end{figure}

The power change of storage exibits a similar pattern: the power change of storage is smooth before $t$=6s and oscillates after $t$=6s, this is because energy in storage nearly touches the upper limits, and then storage can not charge power as before. This makes that power balance can not be maintained and the system lose the frequency stability.
 
\begin{figure}[!tbh]
	\centering
	\includegraphics[width=3.5in]{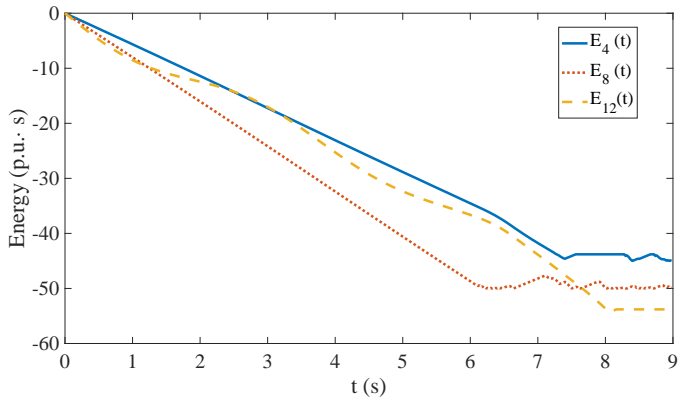}
	\caption{Energy change of storage.}
	\label{nt4}
\end{figure}

\section{conclusion}

In this work, authors give a framework to control virtual inertia and power from storage to regulate the system frequency, Model predictive control is employed to undertake this task, because model predictive control can be implemented for the different system models. For the 2-bus system, centralized MPC is employed, and for the 12-bus system, distributed MPC is employed. 

For the future work, authors want to give a optimization framework to secure the frequency stability of power system where the transmission grid and distribution grid are operated together.

%

\vskip 1cm


%

\appendices
\section{Parameters for Simulation}

\begin{table}[!tbh]
	\renewcommand{\arraystretch}{1.}
	\caption{Inertia and Damping Distribution of Original Power System.}
	\label{tab1}
	\centering
	\begin{tabular}{cc}
		\hline
		\hline
		\textbf{Bus. No.}  &  \textbf{Inertia (s) / Damping (p.u.)}\\
		\hline
		1, 2  &  15/3 \\
		\hline
		5, 6 & 20/4\\
		\hline
		9, 10& 10/2\\
		\hline
		3, 7, 11  & 1/0.1\\
		\hline
		\hline
	\end{tabular}
\end{table}

\begin{table}[!tbh]
	\renewcommand{\arraystretch}{1.}
	\caption{Power Flow Condition.}
	\label{tab22}
	\centering
	\begin{tabular}{c|ccccccc}
		\hline
		\hline
		Gen  &  1  &  2  &  5 & 6  & 9 & 10\\
		\hline
		$P$ (MW)  & 138& 1050 & 719 & 350 & 700 & 700\\
		\hline
		Load  &  3  & 4  &  7 & 8  & 11 & 12\\
		\hline
		$P$ (MW)  &  400& 567 & 490 & 800 & 400 & 1000\\
		\hline
		\hline
	\end{tabular}
\end{table}

\begin{table}[!tbh]
	\scriptsize
	\renewcommand{\arraystretch}{1.}
	\caption{Initial parameters for simulation.}
	\label{tn1}
	\centering
	\begin{tabular}{cc|cc}
		\hline
		\hline
		\textbf{Parameter}  &  \textbf{Value}& \textbf{Parameter}&\textbf{Value}\\
		\hline

		\hline
		$\delta_{1}(0)$ & -0.1931 & $\omega_{1}(0)$ & 0 \\
		\hline
		$\delta_{2}(0)$ & -0.0452 & $\omega_{2}(0)$ & 0 \\
		\hline
		$\delta_{3}(0)$ & -0.2552 &$\omega_{3}(0)$ & 0 \\
		\hline
		$\delta_{4}(0)$ & -0.3340 & $\omega_{4}(0)$ & 0 \\
		\hline
		$\delta_{5}(0)$ & -0.1146 & $\omega_{5}(0)$ & 0 \\
		\hline
		$\delta_{6}(0)$ & -0.3681 & $\omega_{6}(0)$ & 0 \\
		\hline
		$\delta_{7}(0)$ & -0.4381 & $\omega_{7}(0)$ & 0 \\
		\hline
		$\delta_{8}(0)$ & -0.4960 & $\omega_{8}(0)$ & 0 \\
		\hline
		$\delta_{9}(0)$ & 0 &  $\omega_{9}(0)$  & 0 \\
		\hline
		$\delta_{10}(0)$ & -0.1750 &$\omega_{10}(0)$ & 0 \\
		\hline
		$\delta_{11}(0)$ & -0.3150 & $\omega_{11}(0)$ & 0  \\
		\hline
		$\delta_{12}(0)$ & -0.4150 & $\omega_{12}(t)$ & 0  \\
		\hline
		$D_{e,4}$  & 0.1 p.u.  & $D_{e,8}$  &  0.1 p.u. \\
			\hline
		$D_{e,4}$  & 0.1 p.u. &  $M_{e,4}(t)$ &  [4s, 10s] \\
		\hline 
		$M_{e,8}(t)$  & [4s, 10s]  &  $M_{e,12}(t)$ &  [4s, 10s] \\
		\hline
		\hline
	\end{tabular}
\end{table}


\ifCLASSOPTIONcaptionsoff
  \newpage
\fi
\newpage



\bibliographystyle{IEEEtran}
\bibliography{journal}

\begin{thebibliography}{10}
\providecommand{\url}[1]{#1}
\csname url@samestyle\endcsname
\providecommand{\newblock}{\relax}
\providecommand{\bibinfo}[2]{#2}
\providecommand{\BIBentrySTDinterwordspacing}{\spaceskip=0pt\relax}
\providecommand{\BIBentryALTinterwordstretchfactor}{4}
\providecommand{\BIBentryALTinterwordspacing}{\spaceskip=\fontdimen2\font plus
\BIBentryALTinterwordstretchfactor\fontdimen3\font minus
  \fontdimen4\font\relax}
\providecommand{\BIBforeignlanguage}[2]{{%
\expandafter\ifx\csname l@#1\endcsname\relax
\typeout{** WARNING: IEEEtran.bst: No hyphenation pattern has been}%
\typeout{** loaded for the language `#1'. Using the pattern for}%
\typeout{** the default language instead.}%
\else
\language=\csname l@#1\endcsname
\fi
#2}}
\providecommand{\BIBdecl}{\relax}
\BIBdecl

\bibitem{ayodele2015mitigation}
T.~Ayodele and A.~Ogunjuyigbe, ``Mitigation of wind power intermittency:
  Storage technology approach,'' \emph{Renewable and Sustainable Energy
  Reviews}, vol.~44, pp. 447--456, 2015.

\bibitem{ulbig2014impact}
A.~Ulbig, T.~S. Borsche, and G.~Andersson, ``Impact of low rotational inertia
  on power system stability and operation,'' \emph{IFAC Proceedings Volumes},
  vol.~47, no.~3, pp. 7290--7297, 2014.

\bibitem{fang2018inertia}
J.~Fang, H.~Li, Y.~Tang, and F.~Blaabjerg, ``On the inertia of future
  more-electronics power systems,'' \emph{IEEE Journal of Emerging and Selected
  Topics in Power Electronics}, 2018.

\bibitem{zzz3}
B.~Delpech, ``Using ner 300 and the energy recovery plan to renew the energy
  sector,'' European Association for Storage of Energy, Tech. Rep., 2013.

\bibitem{hsieh2012frequency}
E.~Hsieh and R.~Johnson, ``Frequency response from autonomous battery energy
  storage,'' in \emph{Cigr{\'e} Grid of the Future Symposium}, 2012, pp. 1--7.

\bibitem{delille2012dynamic}
G.~Delille, B.~Francois, and G.~Malarange, ``Dynamic frequency control support
  by energy storage to reduce the impact of wind and solar generation on
  isolated power system's inertia,'' \emph{IEEE Transactions on sustainable
  energy}, vol.~3, no.~4, pp. 931--939, 2012.

\bibitem{wen2016frequency}
Y.~Wen, W.~Li, G.~Huang, and X.~Liu, ``Frequency dynamics constrained unit
  commitment with battery energy storage,'' \emph{IEEE Transactions on Power
  Systems}, vol.~31, no.~6, pp. 5115--5125, 2016.

\bibitem{wen2015enhanced}
Y.~Wen, C.~Guo, D.~S. Kirschen, and S.~Dong, ``Enhanced security-constrained
  opf with distributed battery energy storage,'' \emph{IEEE Transactions on
  Power Systems}, vol.~30, no.~1, pp. 98--108, 2015.

\bibitem{kerdphol2018demonstration}
T.~Kerdphol, F.~S. Rahman, V.~Phunpeng, M.~Watanabe, and Y.~Mitani,
  ``Demonstration of virtual inertia emulation using energy storage systems to
  support community-based high renewable energy penetration,'' in \emph{2018
  IEEE Global Humanitarian Technology Conference (GHTC)}.\hskip 1em plus 0.5em
  minus 0.4em\relax IEEE, 2018, pp. 1--7.

\bibitem{tielens2016relevance}
P.~Tielens and D.~Van~Hertem, ``The relevance of inertia in power systems,''
  \emph{Renewable and Sustainable Energy Reviews}, vol.~55, pp. 999--1009,
  2016.

\bibitem{kiaei2018tube}
I.~Kiaei and S.~Lotfifard, ``Tube-based model predictive control of energy
  storage systems for enhancing transient stability of power systems,''
  \emph{IEEE Transactions on Smart Grid}, vol.~9, no.~6, pp. 6438--6447, 2018.

\bibitem{rostami2018scalable}
M.~Rostami and S.~Lotfifard, ``Scalable coordinated control of energy storage
  systems for enhancing power system angle stability,'' \emph{IEEE Transactions
  on Sustainable Energy}, vol.~9, no.~2, pp. 763--770, 2018.

\bibitem{adrees2016study}
A.~Adrees and J.~V. Milanovic, ``Study of frequency response in power system
  with renewable generation and energy storage,'' in \emph{Power Systems
  Computation Conference (PSCC), 2016}.\hskip 1em plus 0.5em minus 0.4em\relax
  IEEE, 2016, pp. 1--7.

\bibitem{guggilam2018optimizing}
S.~Guggilam, C.~Zhao, E.~Dall'Anese, C.~Chen, and S.~Dhople, ``Optimizing der
  participation in inertial and primary-frequency response,'' \emph{IEEE
  Transactions on Power Systems}, 2018.

\bibitem{yan2019controlling}
S.~Yan, ``Controlling time-variant virtual inertia from storage by dynamic
  programming and propt,'' \emph{arXiv preprint arXiv:1903.03790}, 2019.

\bibitem{messac2015optimization}
A.~Messac, \emph{Optimization in practice with MATLAB{\textregistered}: for
  engineering students and professionals}.\hskip 1em plus 0.5em minus
  0.4em\relax Cambridge University Press, 2015.

\bibitem{lei2009review}
M.~Lei, L.~Shiyan, J.~Chuanwen, L.~Hongling, and Z.~Yan, ``A review on the
  forecasting of wind speed and generated power,'' \emph{Renewable and
  Sustainable Energy Reviews}, vol.~13, no.~4, pp. 915--920, 2009.

\bibitem{senjyu2002one}
T.~Senjyu, H.~Takara, K.~Uezato, and T.~Funabashi, ``One-hour-ahead load
  forecasting using neural network,'' \emph{IEEE Transactions on power
  systems}, vol.~17, no.~1, pp. 113--118, 2002.

\bibitem{rawlings2009model}
J.~B. Rawlings and D.~Q. Mayne, \emph{Model predictive control: Theory and
  design}.\hskip 1em plus 0.5em minus 0.4em\relax Nob Hill Pub. Madison,
  Wisconsin, 2009.

\bibitem{camponogara2002distributed}
E.~Camponogara, D.~Jia, B.~H. Krogh, and S.~Talukdar, ``Distributed model
  predictive control,'' \emph{IEEE control systems magazine}, vol.~22, no.~1,
  pp. 44--52, 2002.

\bibitem{venkat2008distributed}
A.~N. Venkat, I.~A. Hiskens, J.~B. Rawlings, and S.~J. Wright, ``Distributed
  mpc strategies with application to power system automatic generation
  control,'' \emph{IEEE transactions on control systems technology}, vol.~16,
  no.~6, pp. 1192--1206, 2008.

\bibitem{bergen1981structure}
A.~R. Bergen and D.~J. Hill, ``A structure preserving model for power system
  stability analysis,'' \emph{IEEE Transactions on Power Apparatus and
  Systems}, no.~1, pp. 25--35, 1981.

\bibitem{bayer2006discretization}
C.~Bayer, ``Discretization of sdes: Euler methods and beyond.''\hskip 1em plus
  0.5em minus 0.4em\relax At PRisMa 2006 Workshop, 2006.

\bibitem{christofides2013distributed}
P.~D. Christofides, R.~Scattolini, D.~M. de~la Pena, and J.~Liu, ``Distributed
  model predictive control: A tutorial review and future research directions,''
  \emph{Computers \& Chemical Engineering}, vol.~51, pp. 21--41, 2013.

\bibitem{morstyn2018model}
T.~Morstyn, B.~Hredzak, R.~P. Aguilera, and V.~G. Agelidis, ``Model predictive
  control for distributed microgrid battery energy storage systems,''
  \emph{IEEE Transactions on Control Systems Technology}, vol.~26, no.~3, pp.
  1107--1114, 2018.

\bibitem{hans2019hierarchical}
C.~A. Hans, P.~Braun, J.~Raisch, L.~Gr{\"u}ne, and C.~Reincke-Collon,
  ``Hierarchical distributed model predictive control of interconnected
  microgrids,'' \emph{IEEE Transactions on Sustainable Energy}, vol.~10, no.~1,
  pp. 407--416, 2019.

\bibitem{chang2014multi}
T.-H. Chang, M.~Hong, and X.~Wang, ``Multi-agent distributed optimization via
  inexact consensus admm,'' \emph{IEEE Transactions on Signal Processing},
  vol.~63, no.~2, pp. 482--497, 2014.

\bibitem{chang2016proximal}
T.-H. Chang, ``A proximal dual consensus admm method for multi-agent
  constrained optimization,'' \emph{IEEE Transactions on Signal Processing},
  vol.~64, no.~14, pp. 3719--3734, 2016.

\bibitem{kundur1994power}
P.~Kundur, N.~J. Balu, and M.~G. Lauby, \emph{Power system stability and
  control}.\hskip 1em plus 0.5em minus 0.4em\relax McGraw-hill New York, 1994,
  vol.~7.

\bibitem{borsche2015effects}
T.~S. Borsche, T.~Liu, and D.~J. Hill, ``Effects of rotational inertia on power
  system damping and frequency transients,'' in \emph{Decision and Control
  (CDC), 2015 IEEE 54th Annual Conference on}.\hskip 1em plus 0.5em minus
  0.4em\relax IEEE, 2015, pp. 5940--5946.

\end{thebibliography}

\end{document}